\documentclass[review]{elsarticle}
\usepackage{amssymb}

\providecommand{\tabularnewline}{\\}

\newcommand{\GeV}{\mbox{GeV}} 
 \newcommand{\TeV}{\mbox{TeV}}
\newcommand{\Br}{\mathrm{Br}}

\newcommand{\Q}{\mathcal{Q}}
\newcommand{\LO}{\mathrm{LO}} 
\newcommand{\OO}{\mathcal{O}}

\journal{Nuclear Physics B}

\begin{document}
\begin{frontmatter}

\title{Hadronic Production of $\chi_{c}$-mesons at LHC}

\author{A.K. Likhoded}

\ead{Anatolii.Likhoded@ihep.ru}

\author{A.V. Luchinsky}

\ead{Alexey.Luchinsky@ihep.ru}

\author{S.V. Poslavsky}

 \address{Institute of High Energy Physics, Protvino, Russia}
\begin{abstract}
Hadronic production of $P$-wave charmonium states $\chi_{cJ}$ in
hadronic interaction is considered. Using experimental results of
CDF and LHCb collaborations we show, that contributions of color-singlet
components are dominant. As for color-octet mechanism,
we show, that contributions from $P$-wave states can also be observed,
while $S$-wave states can be neglected. The best experimental observables
that give information about relative importance of color-singlet and
color-octet components are the ratios $\chi_{c2}/\chi_{c1}$ and $\chi_{c0}/\chi_{c1}$.
\end{abstract}


\end{frontmatter}

\section{Introduction}

Heavy quarkonia production in hadronic experiments is an extremely
interesting task for theoretical and experimental investigation. It
is well known, that at high energies the dominant mechanism for these
processes is gluon-gluon fusion, but at leading order approximation
of perturbation theory such approach cannot describe observed experimentally
production of $\chi_{c1}$ meson and distributions over the transverse
momentum of final quarkonium. This problem can be solved by considering
higher order processes and in the following we show, that in high
$p_{T}$ region is is sufficient to study only NLO approximation and
consider subprocesses $gg\to\chi_{cJ}g$.

The other interesting topic in heavy quarkonia production is the influence
of color octet (CO) components. According to NonRelativistic Quantum
ChromoDynaimcs (NRQCD) \cite{Bodwin:1994jh} the quark-antiquark pair
$c\bar{c}$ in charmonium should not necessary be in the color-singlet
(CS) state. There are also CO components in $\chi_{cJ}$mesons accomplished
with additional gluons. Differential cross sections of production
of such states depends on their quantum numbers, so from analysis
of experimental distributions over different kinematical valiables
(e.g. transverse momentum $p_{T}$) one can determine the relative
contributions of different states into total and differential cross
sections.

The rest of the paper is organized as follows. In the next section
used in our paper partonic subprocesses are briefly discussed. In
section III we present the analysis of experimental data obtained
by CDF and LHCb collaborations and determine contributions of CS and
CO components into $\chi_{c1,2}$ production cross sections, Theoretical
predictions for $\chi_{c0}$ production cross sections and $p_{T}$
dependence of the ratios $\chi_{c0}/\chi_{c1}$, $\chi_{c0}/\chi_{c2}$
are also given in sec.III. Brief analysis of our results is given
in the conclusion.

\section{Partonic Processes}

Our article is devoted to charmonia production in high energy hadronic
experiments at Tevatron and LHC (preliminary discussion of this topic
can be found for example in our previous paper \cite{Likhoded:2012hw}).
It is well known, that main contribution at these conditions comes
from gluon-gluon fusion

\begin{eqnarray}
gg & \to & \left(c\bar{c}\right),\label{proc:LO}
\end{eqnarray}
where quark-antiquark pair hadronizes into color-singlet charmonium
meson $\Q$. The cross section of its hadronic production can be written
in the following form

\begin{eqnarray}
\sigma_{\LO} & = & \int_{0}^{1}dx_{1}dx_{2}f_{g}\left(x_{1}\right)f_{g}\left(x_{2}\right)\hat{\sigma}_{\LO}\left(gg\to\Q\right),\label{eq:LO}
\end{eqnarray}
where $x_{1,2}$ are momentum fractions of the incoming partons, $f_{g}(x_{1,2})$
are distribution functions of these partons in initial hadrons, and
$\hat{\sigma}(gg\to\Q)$ is the cross sections of the hard subprocess
(\ref{proc:LO}).

There are however some drawbacks in such approach. First of all, in
collinear approximation, when transverse motion of initial gluons
is neglected, expression (\ref{eq:LO}) cannot describe the distributions
over the transverse momentum of final charmonium. Moreover, in CS
approximation only mesons with positive charge parity ($\eta_{c}$,
$\chi_{cJ}$) can be produced in reaction (\ref{proc:LO}) and the
case of axial charmonium is excluded by Landau-Yang theorem \cite{Landau:1948kw,Yang:1950rg},
that forbids its production in $gg\to\chi_{c1}$ with two massless
gluons interaction. It is clear, that these results contradict dramatically
existing data, since both transverse momentum distributions and $\chi_{c1}$
meson production were observed experimentally. The cross section of
th latter process is even larger than the cross section of tensor
charmonium production.

\begin{figure}
\begin{centering}
\includegraphics[width=\textwidth]{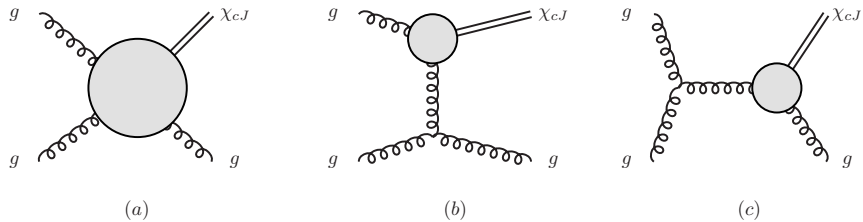}
\par\end{centering}

\caption{Diagrams of the color-singlet $gg\to\chi_{cJ}g$ subprocess\label{diags:CS}}
\end{figure}

These problems can be solved in NLO approximation, when subprocesses

\begin{eqnarray}
gg & \to & \chi_{cJ}g,\label{proc:NLO}
\end{eqnarray}
are considered. Typical diagrams for these reactions are shown in
Fig.\ref{diags:CS}. For the first time they were studied in paper
\cite{Kartvelishvili:1978id}, later a series of papers devoted to
the same topic followed (see for example \cite{Baier:1983va}). It
is clear, that due to presence of final state gluon in (\ref{proc:NLO})
final charmonium has non-vanishing transverse momentum even in collinear
approximation. As for $\chi_{c1}$ meson production, it is allowed
since one of the parent gluons is virtual (Fig.\ref{diags:CS}b) and
Landau-Yang theorem does not forbids the vertex $gg^{*}\to\chi_{c1}.$

In addition to CS states one should also taken into account color
octet contributions. The physical $\chi_{cJ}$ meson can be written
as an infinite series \cite{Bodwin:1994jh}
\begin{eqnarray}
\left|\chi_{cJ}\right\rangle  & \sim & \left|R'(0)\right|^{2}\left|c\bar{c}\left[^{3}P_{J}^{[1]}\right]\right\rangle +\left\langle \OO_{S}\right\rangle \left|c\bar{c}\left[^{3}S_{1}^{[8]}\right]\right\rangle +\left\langle \OO_{P}\right\rangle \left|c\overline{c}\left[^{1}P_{1}^{[8]}\right]\right\rangle \label{eq:fock}
\end{eqnarray}
where in parentheses quantum numbers of quark-antiquark state and
its color charge are shown. The coefficients $\left|R'(0)\right|^{2}$,
$\left\langle \OO_{S,P}\right\rangle $ describe the probability for
corresponding component to hadronize into experimentally observed
meson and are usually taken as universal. According to NRQCD higher
terms in expression (\ref{eq:fock}) are suppressed by relative velocity
of quarks in meson, so this series can be safely truncated. In the
following we restrict ourselves to CS and $S$-, $P$-wave CO. In
this approximation the cross section of $\chi_{cJ}$ meson production
in $gg\to\chi_{cJ}g$ reaction is written in the form
\begin{eqnarray}
\frac{d\hat{\sigma}\left(gg\to\chi_{cJ}g\right)}{d\hat{t}} & = & 
\left|R'(0)\right|^{2}\frac{d\hat{\sigma}\left(gg\to c\overline{c}[^{3}P_{J}^{[1]}]g\right)}{d\hat{t}}+
\nonumber \\
& & \frac{\pi}{6}(2J+1)\left\langle \OO_{S}\right\rangle \frac{d\hat{\sigma}\left(gg\to c\overline{c}[^{3}S_{1}^{[8]}]g\right)}{d\hat{t}}+
\nonumber \\
 &  & \frac{\pi}{18}(2J+1)\left\langle \OO_{P}\right\rangle \frac{d\hat{\sigma}\left(gg\to c\overline{c}[^{1}P_{1}^{[8]}]g\right)}{d\hat{t}},
\end{eqnarray}
where we show explicitly the dependence on wave function derivative
at the origin (in CS case) and matrix elements of $S$-, $P$-wave
octet states. The parameters $|R'(0)|$, $\left\langle \OO_{S,P}\right\rangle $
are universal and do not depend on total spin of the final charmonium.

Hard subprocesses $gg\to\Q g$ were already considered in a number
of papers (see for example \cite{Kartvelishvili:1978id,Baier:1983va,Gastmans:1987be,Meijer:2007eb}),
so below we give only short qualitative analysis. In low $p_{T}$
region some of these cross sections diverge (see second row of table
\ref{tab:CorssSections}). Such behavior is caused by $t$-channel
gluon from diagram shown in Fig.\ref{diags:CS}b, that in this region
approaches the mass shell. In order to regularize this divergence
one should consider higher order processes or perform a suitable cut.
In our paper we consider only high $p_{T}$ region, so this singularity
in not crucial. It is interesting to note, however, that the cross
section of $\chi_{c1}$ production is finite over the whole $p_{T}$
domain. The reason is mentioned above Landau-Yang theorem: in low
$p_{T}$ region the vertex $gg\to\chi_{c1}$ vanishes and caused by
gluon propagator divergence is compensated.

\begin{table}
\begin{centering}
\begin{tabular}{|c|c|c||c|c|}
\hline 
 & $^{3}P_{1}^{[1]}$ & $^{3}P_{0,2}^{[1]}$ & $^{1}P_{1}^{[8]}$ & $^{3}S_{1}^{[8]}$\tabularnewline
\hline 
\hline 
$p_{T}\ll M$ & $\sim p_{T}$ & $\sim1/p_{T}$ & $\sim1/p_{T}$ & $\sim p_{T}$\tabularnewline
\hline 
$p_{T}\gg M$  & $\sim1/p_{T}^{5}$ & $\sim1/p_{T}^{5}$ & $\sim1/p_{T}^{5}$ & $\sim1/p_{T}^{3}$\tabularnewline
\hline 
\end{tabular}
\par\end{centering}

\caption{Behavior of partonic cross sections for different processes in low
and high $p_{T}$ regions\label{tab:CorssSections}}
\end{table}

For our analysis we need also the behavior of different partonic cross
sections in high $p_{T}$ region. Corresponding expressions can be
found in the third row of table \ref{tab:CorssSections}. It is clearly
seen, that in this region cross sections of CS and $P$-wave CO cross
sections are proportional, so it is not sufficient to study $p_{T}$
distributions of different $\chi_{cJ}$ mesons separately to determine
relative contributions of CS and CO components. It is necessary to
consider some combined quantity, for example the ratio
\begin{eqnarray}
\hat{r}_{J_{1}J_{2}} & = & \frac{d\hat{\sigma}\left(gg\to\chi_{cJ_{1}}g\right)/dp_{T}}{d\hat{\sigma}\left(gg\to\chi_{cJ_{2}}g\right)/dp_{T}}.\label{eq:hatR}
\end{eqnarray}
If the contribution of $S$-wave CO states is not negligible, they
should dominate in high $p_{T}$ region and the ration (\ref{eq:hatR})
takes the form
\begin{eqnarray}
\hat{r}_{J_{1}J_{2}}\left(p_{T}\gg M\right) & \approx & \frac{2J_{1}+1}{2J_{2}+1}.
\end{eqnarray}
In the opposite case the ratio for different values of final charmonia
spins is equal to
\begin{eqnarray}
\hat{r}_{2,1} & = & \frac{158\,184\left|R'(0)\right|^{2}+295\,595\pi\left\langle \OO_{P}\right\rangle }{474\,552\left|R'(0)\right|^{2}+177\,357\pi\left\langle \OO_{P}\right\rangle },\\
\hat{r}_{0,1} & = & \frac{79\,092\left|R'(0)\right|^{2}+59\,119\pi\left\langle \OO_{P}\right\rangle }{474\,552\left|R'(0)\right|^{2}+177\,357\pi\left\langle \OO_{P}\right\rangle },\\
\hat{r}_{0,2} & = & \frac{79\,092\left|R'(0)\right|^{2}+59\,119\pi\left\langle \OO_{P}\right\rangle }{158\,184\left|R'(0)\right|^{2}+295\,595\pi\left\langle \OO_{P}\right\rangle }.
\end{eqnarray}
From combined analysis of experimental $p_{T}$ distributions of $\chi_{cJ}$
mesons productions separately and their ratios one can determine contributions
of CS and various CO states to these processes.

\section{Hadronic Production and Fit of Matrix Elements}

For comparison with experimental data considered in the previous section
cross sections should be convoluted with gluon distribution functions
in initial hadrons. Sometimes functions that depend explicitly on
the transverse momentum of the parton are used (so called $k_{T}$
factorization), that take into account multiple emisson of soft gluons.
It is clear, however, that in high $p_{T}$ region such processes
will be suppressed by small string coupling constant, so the emission
of one hard gluon can be preferable. In our paper we use the latter
approach.

The cross section of inclusive $\chi_{cJ}$ production in hadronic
interaction can be written \cite{Likhoded:2007fz}in the form similar
to eq.(\ref{eq:LO}):
\begin{eqnarray}
\frac{d\sigma\left(pp\to\chi_{cJ}+X\right)}{dp_{T}} & = & 
\int\frac{d\hat{s}}{s}\frac{d\hat{\sigma}\left(gg\to\chi_{cJ}g\right)}{dp_{T}}\times
\nonumber\\
&&\int dyf_{g}\left(\sqrt{\frac{\hat{s}}{s}}e^{y}\right)f_{g}\left(\sqrt{\frac{\hat{s}}{s}}e^{-y}\right),\\
\frac{d\hat{\sigma}}{dp_{T}} & = & \frac{2\hat{s}p_{T}}{\sqrt{\left(\hat{s}-M^{2}\right)^{2}-4\hat{s}p_{T}^{2}}}\left(\left.\frac{d\hat{\sigma}}{d\hat{t}}\right|_{\hat{t}=\hat{t}_{1}}+\left.\frac{d\hat{\sigma}}{d\hat{t}}\right|_{\hat{t}=\hat{t}_{2}}\right),
\end{eqnarray}
where $\hat{s},$$\hat{t}$, $\hat{u}$ are the Mandelstam variables
of the partonic subprocess,
\begin{eqnarray}
p_{T} & = & \sqrt{\frac{\hat{t}\hat{u}}{\hat{s}}}
\end{eqnarray}
is the transverse momentum if final charmonium (in collinear approximation
this expression is valid both for partonic and hadronic reactions),
and the following notations were introduced:
\begin{eqnarray}
\hat{t}_{1,2}\left(p_{T}\right) & = & \frac{1}{2}\left\{ M^{2}-\hat{s}\pm\sqrt{\left(\hat{s}-M^{2}\right)^{2}-4\hat{s}p_{T}^{2}}\right\} .
\end{eqnarray}
Partonic cross sections and distribution functions that enter these
expressions depend strongly on factorization scale $\mu^{2}$. In
order to study the dependence of final results on the choice of this
scale we use the following values: $\mu^{2}=M^{2}$, $\mu^{2}=m_{T}^{2}=p_{T}^{2}+M^{2}$,
$\mu^{2}=2m_{T}^{2}$ and $\mu^{2}=m_{T}^{2}/2$. Gluon distribution
functions were taken in CTEQ6 parameterization. 

For determination of CS and CO parameters $\left|R'(0)\right|^{2}$,
$\left\langle \OO_{S,P}\right\rangle $ we use exprerimental data
obtained by CDF \cite{Abe:1997yz,Abulencia:2007bra} and LHCb \cite{LHCb:2012ac}
collaborations. In paper \cite{Abe:1997yz} transverse momentum distribution
of $J/\psi$-meson production cross section in radiative decays $\chi_{cJ}\to J/\psi\gamma$
at $\sqrt{s}=1.8\,\TeV$, $\left|\eta\right|<0.6$ is presented:
\begin{eqnarray}
\frac{d\sigma\left(pp\to J/\psi+X\right)}{dp_{T}} & = & \sum_{J=0}^{2}\Br\left[\chi_{cJ}\to J/\psi\gamma\right]\frac{d\sigma\left(pp\to\chi_{cJ}+X\right)}{dp_{T}}\label{eq:SigmaPsi}
\end{eqnarray}
Using discussed above theoretical predictions of CS and CO $\chi_{cJ}$
meson differential cross sections and experimental values of radiative
decay branching fractions \cite{Beringer:1900zz} one can determine
parameters $\left|R'(0)\right|^{2}$, $\left\langle \OO_{S}\right\rangle $,
$\left\langle \OO_{P}\right\rangle $ from eq.(\ref{eq:fock}). Due
to small value of the branching fraction of the decay $\chi_{c0}\to J/\psi\gamma$
the scalar charmonium can be excluded from expression (\ref{eq:SigmaPsi}).

We have already stressed above that in high $p_{T}$ region up to
constant factor differential of CS and $P$-wave CO cross sections
coincide, so one should use some other variable to separate these
components. One of such observables is the ratio
\begin{eqnarray}
r_{J_{1}J_{2}}\left(p_{T}\right) & = & \frac{d\sigma\left(pp\to\chi_{cJ_{1}}+X\right)/dp_{T}}{d\sigma\left(pp\to\chi_{cJ_{2}}+X\right)/dp_{T}}.\label{eq:PPratio}
\end{eqnarray}
In high $p_{T}$ region, where hard cross sections of $gg\to\chi_{cJ1}g$
and $gg\to\chi_{cJ2}g$ are almost proportional to each other, partonic
distribution functions in this ratio cancel and it becomes equal to
the ratio of hard cross sections (\ref{eq:hatR}). It should be noted,
that this cancellation is universal and doest not depend on experimental
cutoffs.

\begin{table}
\begin{centering}
\begin{tabular}{|c|c|c|c|}
\hline 
$\mu^{2}$ & $\left|R'(0)\right|^{2},\,\GeV^{5}$ & $\left\langle \OO_{S}\right\rangle ,\,\GeV^{3}$ & $\left\langle \OO_{P}\right\rangle ,\,\GeV^{5}$\tabularnewline
\hline 
\hline 
$M^{2}$ & $0.22$ & $1.9\times10^{-9}$ & $0.029$\tabularnewline
\hline 
$m_{T}^{2}/2$ & $0.20$ & $0$ & $0.026$\tabularnewline
\hline 
$m_{T}^{2}$ & $0.19$ & $4.8\times10^{-11}$ & $0.022$\tabularnewline
\hline 
$2m_{T}^{2}$ & $0.19$ & $4.7\times10^{-9}$ & $0.019$\tabularnewline
\hline 
\end{tabular}
\par\end{centering}

\caption{CS and CO model parameters for different values of the scale $\mu^{2}$\label{tab:fit}}
\end{table}

\begin{figure}
\begin{centering}
\includegraphics[width=\textwidth]{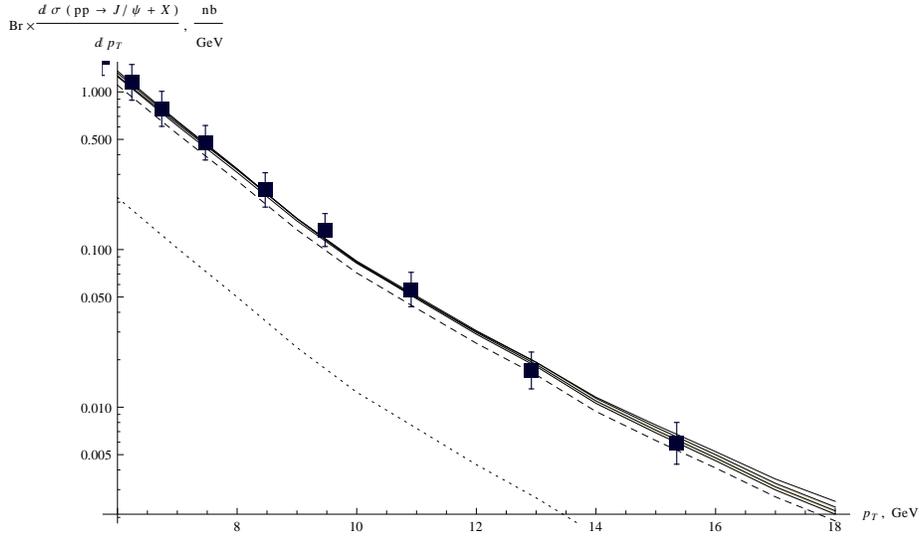}
\par\end{centering}

\caption{Transverse momentum distribution of $J/\psi$ production in radiative
$\chi_{cJ}$ decays at CDF in comparison with experimental data \cite{Abe:1997yz}.
Solid, dashes and dotted lines correspond to total $pp\to\chi_{cJ}+X\to J/\psi+X$
cross section, CS contributions and $P$-wave CO contributions respectively.\label{fig:sigma}}
\end{figure}

The results of our fit for different values of the scale $\mu^{2}$
are presented in table \ref{tab:fit} and $p_{T}$ distribution of
$J/\psi$ production cross section in comparison with experimental
data is shown in fig.\ref{fig:sigma}. It is clear, that results of
our model are in excellent agreement with experiment and contribution
of CS states dominate. As for CO stare, our results show, that contribution
of $S$-wave components can be safely neglected, while $P$-wave CO
contributions are small but visible. This conclusion contradicts NRQCD
scaling rules, that state that CS and $S$-wave CO states should give
similar contributions, while $P$-wave CO sates should be suppressed.

\begin{figure}
\begin{centering}
\includegraphics[width=\textwidth]{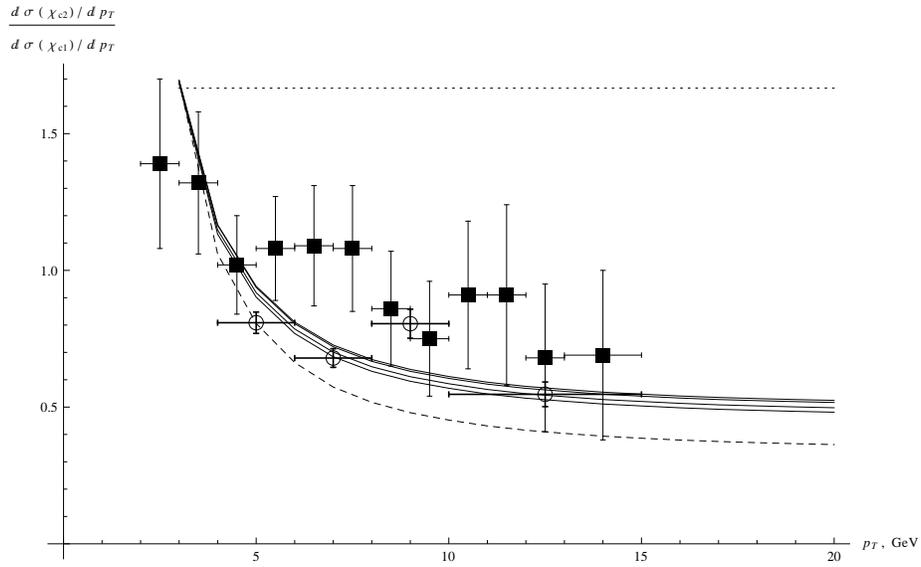}
\par\end{centering}

\caption{Theoretical results of the ratio $\chi_{c2}/\chi_{c1}$ in comparison
with experimental data from \cite{Abulencia:2007bra} ($\circ$) and
\cite{LHCb:2012ac} ($\blacksquare$). Solid lines correspond to parameter
values presented in table \ref{tab:fit}, while dashed and dotted
lines show predictions of our model with only CS and CO contributions
taken into account.\label{fig:ratio}}
\end{figure}

In figure \ref{fig:ratio} we compare theoretical results for $\chi_{c2}/\chi_{c1}$
ratio (solid lines) with experimental data from \cite{LHCb:2012ac,Abulencia:2007bra}
(dots with error bars). Earlier we have said that this ratio can be
used to separate contributions from CS and $P$-wave CO components.
Dashed and dotted lines in this figure show theoretical predictions
of this ratio with only CS or $P$-wave CO contributions taken into
account. It can be clearly seen, that CS mechanism is dominant, but
some CO contribution is also required.

\begin{figure}
\begin{centering}
\includegraphics[scale=0.5]{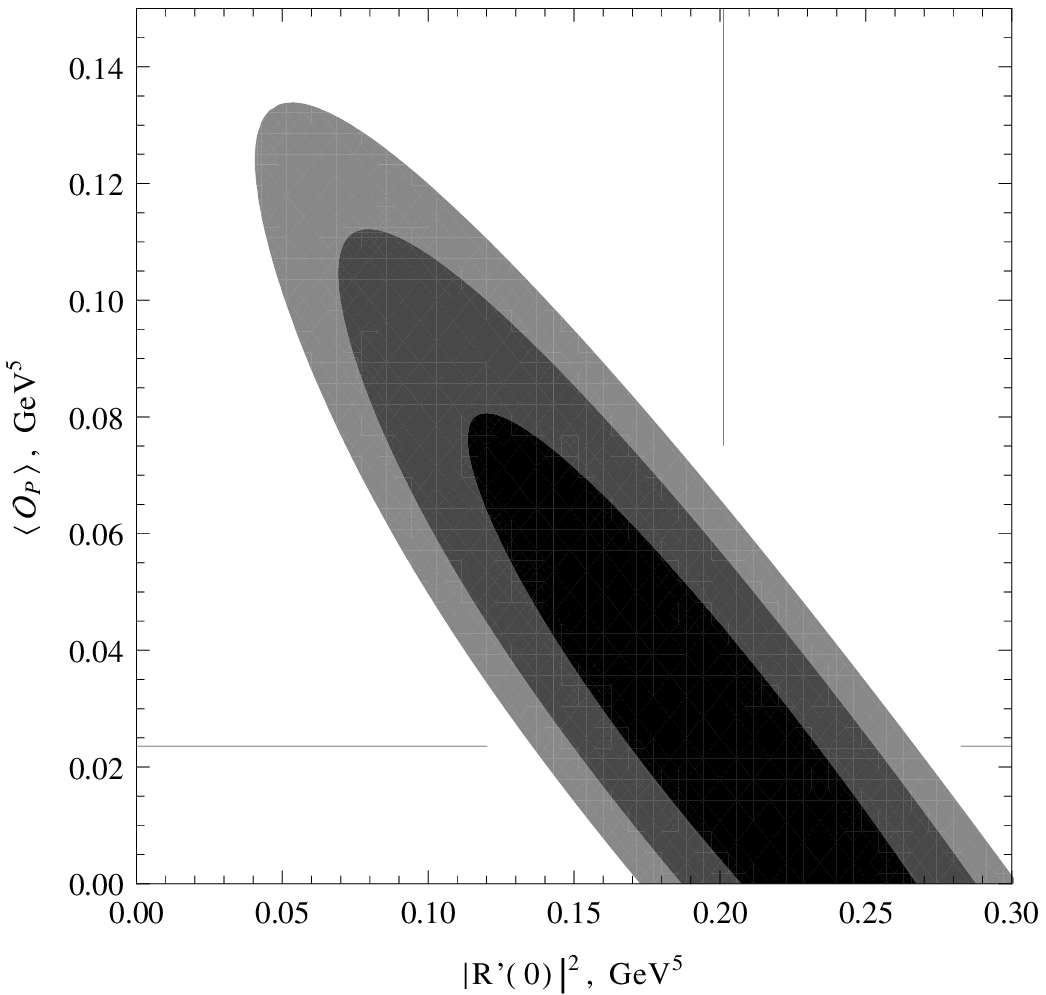}
\par\end{centering}

\caption{Allowed region of parameters $\left|R'(0)\right|^{2}$ and $\left\langle \OO_{P}\right\rangle $\label{fig:chi2}}
\end{figure}

It should be noted, that presented in table \ref{tab:fit} values
of color-singlet parameter $\left|R'(0)\right|^{2}$ are higher than
phenomenological value $\left|R'(0)\right|^{2}\approx0.08\,\GeV^{5}$,
determined from experimental hadronic width of $\chi_{c2}$ meson
\cite{Olsson:1984im}

\begin{eqnarray*}
\Gamma\left(\chi_{c2}\to\mbox{hadrons}\right) & \approx & \Gamma(\chi_{c2}\to2g)=\frac{128}{5}\frac{\alpha_{s}^{2}}{M^{4}}\left|R'(0)\right|^{2}
\end{eqnarray*}
and predictions of different potential models \cite{Munz:1996hb,Ebert:2003mu,Anisovich:2005jp,Wang:2009er,Li:2009nr,Hwang:2010iq}.
One should take into account, that presented in table \ref{tab:fit}
results are strongly correlated. It can be seen clearly from figure
\ref{fig:chi2}, where we show the allowed region of parameters $\left|R'(0)\right|^{2}$,
$\left\langle \OO_{P}\right\rangle $, where the error $\chi^{2}/DOF$
is increased by one unit maximum. The very use of potential model
predictions and $\chi_{c2}$ decay width for charmonium production
at high energies is also rather questionable. From double charmonia
production in exclusive electron-positron annihilation \cite{Abe:2002rb,Braguta:2008qe,Braguta:2008hs}
we know, the with the increase of the interaction energy the width
of the momentum distributions of heavy quarks in quarkonia also increases.
In coordinate space it corresponds to the incrase of the charmonium
wave function and its derivative at the origin. From Fig.\ref{fig:chi2}
it is clear, that such modification of $\left|R'(0)\right|^{2}$ leads
to decrease of the parameter $\left\langle \OO_{P}\right\rangle $
and the contributions from color octet states.

\begin{figure}
\begin{centering}
\includegraphics[width=\textwidth]{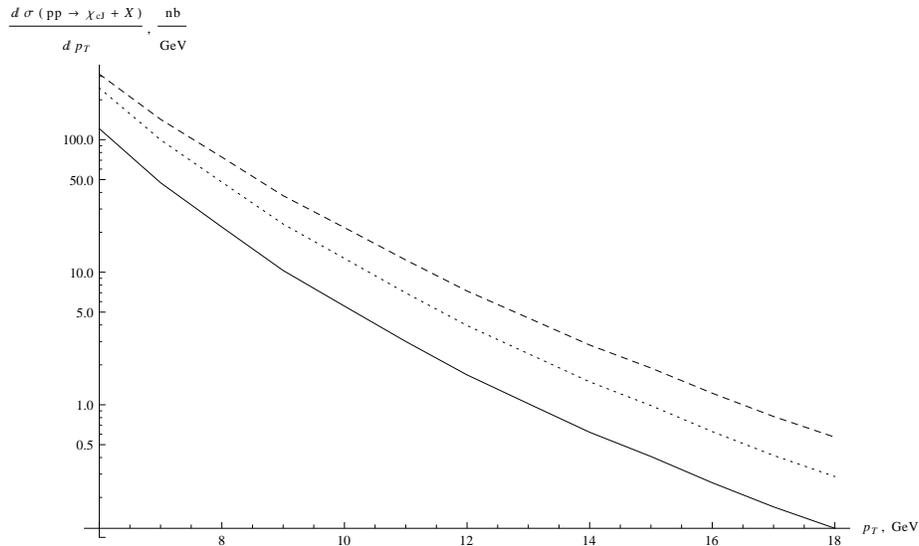}
\par\end{centering}

\caption{Transverse momentum dependence of $\chi_{c0}$, $\chi_{c1}$, and
$\chi_{c2}$ mesons at LHCb (solid, dashed and dotted lines respectively)\label{fig:chiCLHCb}}
\end{figure}

\begin{figure}
\begin{centering}
\includegraphics[width=\textwidth]{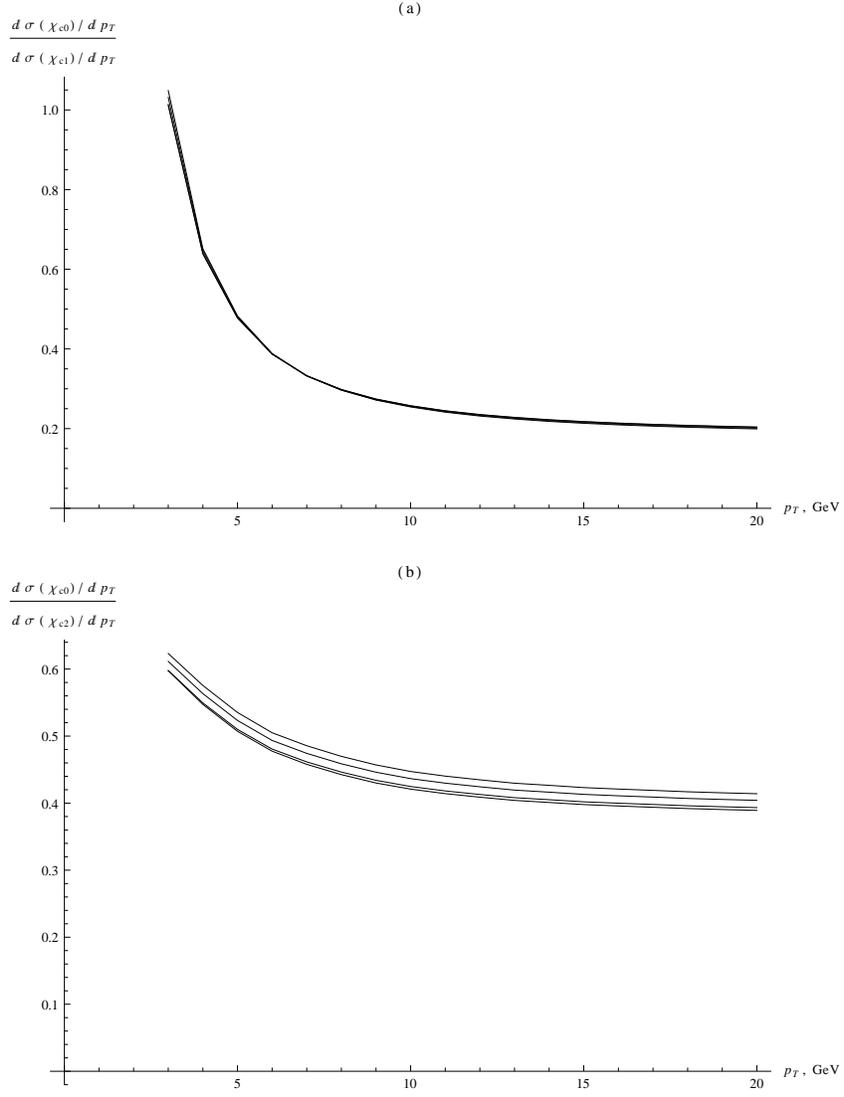}
\par\end{centering}

\caption{The ratios $\chi_{c0}/\chi_{c1}$ (top panel) and $\chi_{c0}/\chi_{c2}$
(bottom panel) at LHCb \label{fig:RatioChiC0}}
\end{figure}

To remove this error one can measure with better accuracy cross sections
of $\chi_{c1,2}$ mesons production and their ratios in various experimental
conditions. The other experiment, that can shed light onto this question
is the observation of $\chi_{c0}$ meson. The branching fraction of
its radiative decay is small, so this task looks very difficult, but
nevertheless possible. In Figs.\ref{fig:chiCLHCb}, \ref{fig:RatioChiC0}
we show theoretical predictions of $\chi_{c0,1,2}$ meson production
cross sections and their ratios at LHCb.

\section{Conclusion}

The paper is devoted to inclusive $P$-wave charmonia production in
hight energy hadronic experiments.

Using existing experimental data presented by collaborationsCDF \cite{Abe:1997yz,Abe:2002rb}
and LHCb \cite{LHCb:2012ac} we determined the cross sections of color
singlet and color octet $\chi_{cJ}$ mesons. Our analysis show, that
contributions of color singlet components are dominant, while $P$-wave
color octet components are strongly suppressed. As for $S$-wave color
octet components, we found, that their contributions can safely be
neglected completely. We also present theoretical predictions for
$\chi_{c1,2}$ production cross sections and transverse momentum distributions
at LHCb and discuss in in details processes of scalar charmonium production
$\chi_{c0}$.

The authors would like to thank A. Novoselov, I. Belyaev and E. Tournefier
for fruitful discussions. The work was financially supported by Russian
Foundation for Basic Research (grant \#10-00061a), the grant of the
president of Russian Federation (grant \#MK-3513.2012.2), and FRRC
grant.

\bibliographystyle{model1a-num-names}
\bibliography{litr}

\end{document}